\documentclass{revtex4}
\usepackage{graphicx}
\usepackage{amsmath}
\usepackage{amsfonts}
\usepackage{braket}
\usepackage{leftidx}
\usepackage{caption}
\usepackage{relsize}
\usepackage{amssymb}
\newcommand{\Conv}{\mathop{\scalebox{2.3}{\raisebox{-0.0ex}{$\circledast$}}}}%
\usepackage{tikz}

\usepackage{mathtools}

\DeclarePairedDelimiter\ceil{\lceil}{\rceil}

\newcommand{\emma}[1]{{#1}}

\begin{document}

\title{Configurational and energy landscape in one-dimensional Coulomb
  systems}

\author{Lucas Varela} \author{Gabriel T\'ellez}
\affiliation{Departamento de F\'{\i}sica, Universidad de los Andes,
  Bogot\'a, Colombia}

\author{Emmanuel Trizac}
\affiliation{LPTMS, CNRS, Univ. Paris-Sud, Universit\'e Paris-Saclay, 91405 Orsay, France}

\begin{abstract}
We study a one dimensional Coulomb system, where two charged colloids are neutralized by a
collection of point counterions, with global neutrality. Temperature being given, two situations are addressed: the colloids
are either kept at fixed positions (canonical ensemble), or the force
acting on the colloids is fixed (isobaric-isothermal ensemble). The corresponding partition functions are worked out
exactly, in view of determining which arrangement of counterions is optimal: how many counterions should be in the
confined segment between the colloids? For the remaining ions outside, is there a left/right symmetry breakdown?
We evidence a cascade of transitions, as system size is varied in the canonical treatment, or as pressure
is increased in the isobaric formulation.
\end{abstract}

\maketitle

\section{Introduction}\label{I}

In condensed matter physics, interactions due to electrostatic
forces are essential. Since matter is made of protons and
electrons, special properties of
materials are ultimately due to the electric and magnetic interactions
at the atomic and molecular scales. Plasmas exhibit as well strong
electromagnetic interactions which are responsible for their
behavior. Their understanding sheds light on the physics at work in the
interior of a star \cite{levin}, or on the conducting properties of
liquid metals such as gallium based alloys, which are used for
industrial purposes. Substances with polar solvents (e.g.~colloids,
polymers and membranes) give another example of systems where
including electric interactions is paramount to describe
thermodynamic properties \cite{messina}.

Coulomb systems are the ensembles which model the interaction between
charges with the Coulomb potential energy. In three dimensions (3d), this
potential reads
\begin{equation}\label{3dcol}
U_{C_{3d}}( \textbf{r}_1, \textbf{r}_2 ) =  \frac{-q_1q_2}{|\textbf{r}_1-\textbf{r}_2|}\,.
\end{equation}
Yet, an analytic expression for a
partition function of a Coulomb system is impossible to obtain. Mean
field theories provide an approach to this problem but fail when the
electric correlations are important, which is the case for many
systems of interest in soft condensed matter \cite{levin,messina}. The two
most important phenomena they fail to account for are charge
reversal (overcharging) and like charge attraction. The first occurs
when a colloid (large ion) is screened by enough counterions (small
ions with opposite charge) such that net charge opposes the one of the
colloid. Like charge attraction happens when two colloids of the same
bare charge sign are attracted to each other due to the interactions with
the medium (counterions).

Another powerful method that has proven operational predicting the
behavior of Coulomb systems, especially the qualitative one, is to study
a lower dimension model. In lower dimensions, the electrostatic
potential is easier to manipulate while keeping important features of
the 3d case such as its long range. They can also be mapped to some
tridimensional systems with a translational invariance in one
direction. For example, the two-dimensional (2d) logarithmic potential
given by
\begin{equation}\label{2dcol}
U_{C_{2d}}( \textbf{r}_1, \textbf{r}_2 ) = -q_1q_2  \ln \frac{|\textbf{r}_1-\textbf{r}_2|}{L}
\end{equation}
is used to model the interaction between vortices in quantum fluids
(e.g.~superfluid $\leftidx{^4}{\text{He}}{}$ and
$\leftidx{^3}{\text{He}}{}$ films), two-dimensional crystalline
solids, and XY (classical rotor) magnets \cite{levin}. Another example
is the analogy between the Laughlin trial wave function for the
fractional quantum Hall effect \cite{laughlin} and the Boltzmann
factor for the two dimensional one component
plasma~\cite{ForresterBook}. This analogy has proven fruitful to
understand the properties of both systems. For example,
in~\cite{can2014, can}, the authors considered a special case which
allows an analytic solution. This led them to find that plasma forms a
double layer structure which causes an excess density as the edge of
the leading support is approached from the inside of the
plasma. Furthermore, these two dimensional models have been used to
study important physical phenomena such as charge renormalization and
the Onsager-Manning-Oosawa counterion condensation \cite{manning, TT2006}
which is reviewed in~\cite{royal}.

\begin{figure}[h]
		\centering
		\includegraphics[width=1\linewidth]{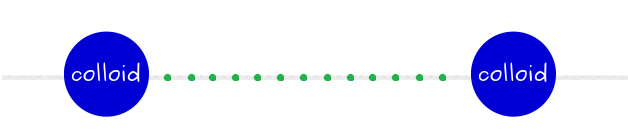}
		\captionof{figure}{Sketch of the model studied in~\cite{tellez}.}
\label{tellezim}
\end{figure}

In one dimensional Coulomb systems, charged particles interact via
the potential
\begin{equation}\label{1dcol}
U_{C_{1d}} (x_1,x_2) =  -q_1q_2|x_1-x_2|
\,.
\end{equation}
For these systems, analytical expressions for the canonical and
isobaric partition functions can be obtained. These were first
investigated by A.~Lenard~\cite{lenard} and S.~Prager~\cite{prager}
independently in 1961. Although a one dimensional model is a
simplification or fictitious model, it gives insights on the
qualitative behavior of the three-dimensional problem. As S.~Prager
remarks in~\cite{prager}: ``It is these (long range) forces which make
the statistical mechanics of plasmas and electrolyte solutions so
extraordinarily difficult to treat. (...)  The one-dimensional plasma,
where this can be done exactly, should thus serve as a useful
testing-ground for approximations developed to treat the
three-dimensional case.''.
The fact that an exact analytical resolution is possible is indeed
of particular interest, for it heralds non mean-field
effects such as overcharging and like charge attraction. These properties
were found in a one dimensional model studied in \cite{tellez}.
It consists of two colloidal
charges separated by a distance $L$ and $N$ neutralizing counterions
bounded by the colloids (Fig.~\ref{tellezim}).  Depending on
whether the colloidal charges are integer multiples of the counterion
charges and the parity of the number of counterions,
it was found that both
overcharging and charge like attraction were possible,

\begin{figure}[h]
	\centering
	\includegraphics[width=1\linewidth]{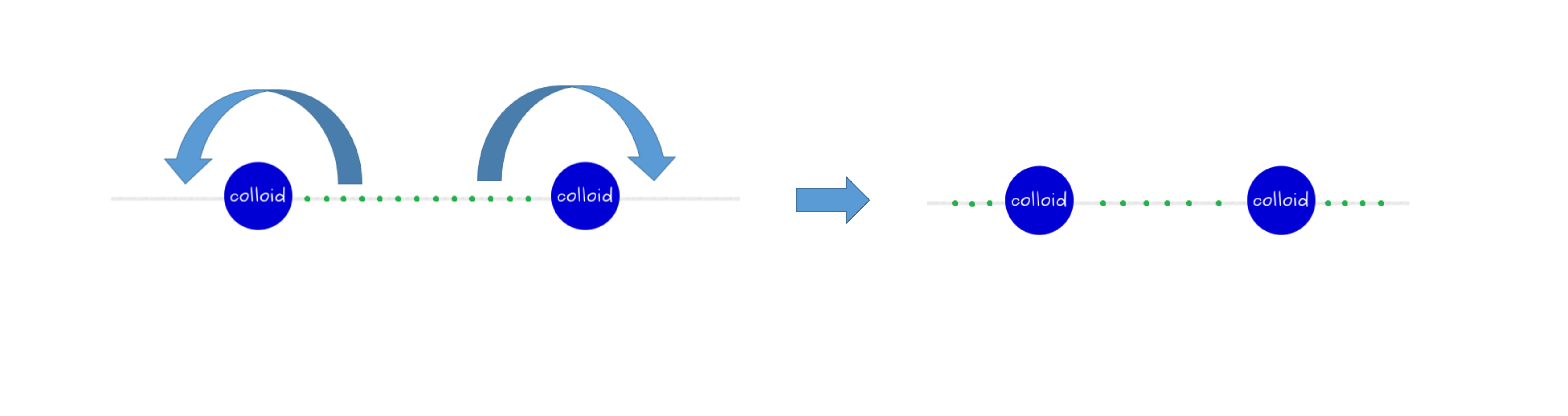}
	\captionof{figure}{Representation of the modified model studied here.}
	\label{colloidim}
\end{figure}

The present work extends the analysis of Ref~\cite{tellez}. We wish here to explore the situation where
a variable number of counterions may become unbounded (Fig.~\ref{colloidim}).
Two types of questions then arise: Which is the optimal configuration, and
do the non mean-field phenomena still occur for the new
configurations? We will answer these questions, providing a complete
thermodynamical solution and interpretation for the canonical and
isobaric ensembles that arise from this modification.

The scheme we use to compute the canonical partition function consists
in rearranging the terms of the Coulomb potential and writing it as a
convolution product of some auxiliary functions. Then, we compute the
Laplace transform of the canonical partition function, which is the
product of the Laplace transforms of each auxiliary function by
virtue of the convolution theorem. By performing the Laplace transform,
we obtain the isobaric partition function which gives us
information about the system in this ensemble. Finally, to obtain the
canonical partition function, we use the inversion formula for Laplace
transform which is computed by using the residue theorem.


The outline is as follows. In section~\ref{II}, following
S.~Prager~\cite{prager} and A.~Lenard~\cite{lenard}, we start by
studying the isobaric ensemble, where the partition function and ensuing
quantities are easier to compute and interpret. Special interest is
given to finding the configuration of particles that minimizes the
Gibbs energy. We then calculate in section \ref{III} the canonical
partition function from its isobaric counterpart, performing an
inverse Laplace transform. In each section, the form of the
thermodynamic quantities in each ensemble is analyzed by analytical
and graphical means.

\section{Isobaric Ensemble}\label{II}

Both Lenard~\cite{lenard} and Prager~\cite{prager} noticed that for a
one dimensional plasma with global neutral charge, the potential energy
could be expressed as a sum of the relative distance between
charges. We follow this technique to rewrite the canonical partition
function in a form in which the Laplace transform (isobaric partition
function) is readily obtained.
%

\subsection{Isobaric partition function}

Consider two charges $q$ along a line located at $\widetilde{x} = 0$ and
$\widetilde{x} = \widetilde{L}$. They play the role of the ``colloids''
depicted in Figs.~\ref{tellezim} and \ref{colloidim}.
There are also $N$ counterions of charge $e =
-2q/N$ with positions $\widetilde{x}_i$. The potential energy of this neutral
system is denoted by $\widetilde{U}$. The system is in thermal equilibrium
at a temperature $T$. It is convenient to introduce the dimensionless
quantities $x = \beta e^2 \widetilde{x}$, for the positions, and $U =
\beta \widetilde{U}$, for the potential energy, where $\beta = (k_B
T)^{-1}$ and $k_B$ is Boltzmann constant.  With this notation, the dimensionless potential energy
takes the following form
%
\begin{equation}
\label{adimpot} U(x_1,\dots, x_N) = \frac{N}{2} \sum_{i=1}^{N} \left(|x_i| + |x_i - L|\right) -\sum_{1\leq i < j \leq N} |x_i - x_j|  - \left(\frac{N}{2}\right)^2 L
\end{equation}
where the summands from left to right are due to colloid-counterion,
counterion-counterion and colloid-colloid interaction
respectively.

For the computation of the canonical partition function, it is
convenient to separate the dimensionless potential energy in three
summands. Let $\ell$ and $r$ be the number of unbounded counterions to
the left and right, i.e for $i\leq \ell$, $x_{i} <0$ and for $i>N-r$,
$x_{i}>L$ respectively. Then $U_L$ and $U_R$ are the contributions due
to the unbounded counterions. $U_B$ is due to the counterions bounded
by the colloids. The expression in terms of the separated potentials
is the following
%
\begin{equation}
\label{potdesc} U(x_1,\dots, x_N) = U_L(x_1,\dots, x_{\ell}) + U_B(x_{\ell+1},\dots, x_{N-r}) +U_R(x_{N-r+1},\dots, x_{N})   \,.
\end{equation}
To simplify the expressions for the potentials the particles are ordered as $x_1 < x_2 < \dots < x_N$. Writing the distance between particles as the sum of the distances between first neighbors, the following expressions are obtained
%
\begin{align}
\label{potleft}  U_L(x_1,\dots, x_{\ell}) =& \sum_{k=1}^{\ell} k^2 (x_{k+1} - x_k) \qquad &\text{with the convention } x_{\ell+1} \equiv 0 \\
\label{potbounded}  U_B(x_{\ell+1},\dots, x_{N-r}) =& -\sum_{k=\ell}^{N-r} k(N-k) (x_{k+1} - x_k) + \frac{N(N-4r)L }{4}\qquad &\text{with } x_{\ell}= x_{N-r+1} \equiv 0\\
\label{potright}  U_R(x_{N-r+1},\dots, x_{N}) =& \sum_{k=N-r}^{N} (N-k)^2 (x_{k+1} - x_k) \qquad &\text{with } x_{N-r} \equiv 0
\,.
\end{align}

The expression for the canonical partition function is given by
\begin{equation}
\label{canon}  Z_c(N,L) = \int_{L}^{\infty}  dx_N \int_{L}^{x_N} dx_{N-1}\dots \int_{L}^{x_{N-r+2}}  dx_{N-r+1}  \int_{0}^{L} dx_{N-r}
\int_{0}^{x_{N-r}} \dots  \int_{0}^{x_{\ell+2}}  \int_{x_{\ell-1}}^{0} \dots
\int_{x_1}^{0} dx_2\int_{-\infty}^{0} e^{-U(x_1,\dots, x_N)} dx_1
\,.
\end{equation}
This expression can be split in a product of three terms, in a similar
way as the potential energy with the aid of the Fubini theorem.
\begin{equation}
\label{canonsep}  Z_c(N,L) = \left( \int_{x_{\ell-1}}^{0} \dots   \int_{-\infty}^{0} e^{-U_L} \prod_{j=1}^{\ell}dx_{j} \right)  \left(\int_{0}^{L} \dots  \int_{0}^{x_{\ell+2}} e^{-U_B} \prod_{j=\ell +1}^{N-r}dx_{j} \right)      \left(\int_{x_{N-1}}^{\infty}\dots \int_{L}^{\infty}  e^{-U_R} \prod_{j=N-r+1}^{N}dx_{j} \right)
\,.
\end{equation}
Grouping the integrals of the left, right and bounded positions of the counterions, a product of the form
\begin{equation}
  \label{eq:prodZ}
  Z_c = Z_L Z_B Z_R
\end{equation}
is obtained. To compute each term, the following auxiliary functions are introduced
\begin{align}
\label{g} g_k(x)  \equiv& \;e^{k^2 x} H(-x)\\
\label{f} f_k(x)  \equiv& \;e^{k(N-k) x}  H(x)
\end{align}
where $H(x)$ is the Heaviside step function. With these functions, the
partition function can be recast as a convolution product. The left partition function $Z_L$ is naturally expressed in
terms of the $g_k$ functions. To write the right term $Z_R$ in terms of the
$g_k$ functions, a change of variables is performed $y_k = x_k -
L$. This translation only adds a factor due to the term that does not
cancel out, $x_{N-r+1} = L + y_{N-r+1}$. The results for $Z_L$ and $Z_R$ are
\begin{align}
\label{canonleft}  Z_L(\ell) =& \; \mathcal{L}\left\{ \Conv^{\ell}_{k=1} g_k(-x_1) \right\}(0) = \left(\frac{1}{\ell!}\right)^2\\
\label{canonright}  Z_R(r) =& \; e^{-r^2 L}      \mathcal{L}\left\{ \Conv^{r}_{k=1} g_k(-y_N) \right\}(0) = \left(\frac{1}{r!}\right)^2 e^{-r^2 L}
\end{align}
where $\mathcal{L} \{ f(x) \}(0)$ is the one sided Laplace transform
of $f(x)$ evaluated at 0.

The partition function $Z_B$, corresponding to the counterions bound
in between the two colloids, is essentially the same as for the
configuration of equally charged colloids studied in
\cite{tellez}. There is a subtle difference with~\cite{tellez}: when
writing it as a convolution product, an extra factor $\exp(r(N-r)L)$
appears.
The expression for the partition function $Z_B$ is
\begin{equation}
\label{canonbounded}  Z_B(N,\ell,r,L) = e^{-r(N-r) L -\frac{N(N-4r)}{4}L}      \Conv^{N- r}_{k=\ell} f_k(L)
\end{equation}

Putting together all these results, the canonical partition function is then written as
\begin{equation}
\label{canons}  Z_c(N,L) = \frac{e^{-\frac{N^2}{4}L}}{(\ell!r!)^2}      \Conv^{N- r}_{k=\ell} f_k(L)
\end{equation}
It is convenient to switch to the isobaric partition function which is the Laplace transform of expression (\ref{canons}). This is done using the convolution theorem, obtaining a simple expression in terms of products of the Laplace transforms of the functions $f_k$ evaluated at $P + N^2/4$ due to the exponential factor,
\begin{equation}
\label{isobaric}  Z_p(N,P) = \frac{1}{(\ell!r!)^2}      \prod_{k=\ell}^{N-r} \frac{1}{P + N^2/4 - k(N-k)}\,.
\end{equation}

To analyze the structure of the isobaric partition function, we examine the effects of the parity of $N$ separately. We distinguish the even $N=2n$  and odd $N =2n+1$ cases. It is also convenient to introduce the parameters $M= \max(\ell,r)$ and $m=\min(\ell,r)$. In the even case $N=2n$, depending on the values of $M$ and $n$, we have
\begin{equation}\label{barieven}
Z_p(2n,P) =\begin{dcases}
\frac{1}{(M!m!)^2}      \frac{1}{P}\left( \prod_{k=1}^{n-M}  \frac{1}{P+k^2}   \right)^2 \prod_{k=n+1-M}^{n-m}\frac{1}{P+k^2}     & \text{if \;}  n-M-\frac{1}{2} > 0\\
\frac{1}{(M!m!)^2}  \prod_{k=M-n}^{n-m}  \frac{1}{P+k^2}&  \text{if \;} \frac{1}{2}+M-n>0
\,.
\end{dcases}
\end{equation}
Several important properties can be seen from these equations. When there is an even amount of counterions, second order poles are present for $M<n$. The leading (largest) pole is $P=0$ for $M\leq n$. For $M>n$ the leading pole is $P = -(M-n)^2$.
For any value of $M$ and $m$ the leading pole is always simple.

In the odd case $N=2n+1$, the partition function is
\begin{equation}\label{bariodd}
Z_p(2n+1,P) = \begin{dcases}
\frac{1}{(M!m!)^2}      \left( \prod_{k=0}^{n-M}  \frac{1}{P+(k+\frac{1}{2})^2}   \right)^2 \prod_{k=n+1-M}^{n-m}\frac{1}{P+(k+\frac{1}{2})^2}       & \text{if \;\;}  n-M+\frac{1}{2} > 0 \\
 \frac{1}{(M!m!)^2}\prod_{k=M-n-1}^{n-m}  \frac{1}{P+(k+\frac{1}{2})^2} &  \text{if\;\;}  M-\frac{1}{2}-n>0
\end{dcases}
\end{equation}

For the odd case the second order poles appear for $M \leq n$ and the leading pole is $P=-1/4$ until $M\leq n+1$. For $M> n+1$ the leading pole is $P= -(M-n-1/2)^2$. In the odd case the leading pole has order 2 for $M< n + 1/2$ and for $M> 1/2 + n$ it becomes simple.

\subsection{Equivalent model}

Consider the model studied in \cite{tellez}. It consists of
$N'$ counterions all bounded without chance to escape and two
colloidal charges surrounding them. These colloids have different
charge magnitudes, $Q_1$ and $Q_2$. As considered here,
the system is neutral. After comparing the isobaric partition
function (\ref{isobaric}) with the expression obtained for the
screening of two unequal charges in \cite{tellez}, we found they are
proportional by a factor $(M! m!)^2$
\begin{equation}
\label{equiv1} Z_p(N,P) = \frac{1}{(M!m!)^2}  {\cal Z}_p(N'=N-M-m,P,Q_1,Q_2)
\end{equation}
where
\begin{equation}
\label{equiv} {\cal Z}_p(N',P,Q_1,Q_2) =  \prod_{k=M}^{N-m} \frac{1}{P + N^2/4 - k(N-k)} =
\substack{\mathlarger{\mathlarger{\prod}} \\
k \in \{ -Q_{<}, -Q_{<}+1, \dots,Q_{>}-1,Q_{>}       \}}
\frac{1}{P + k^2}
\,
\end{equation}
is the partition function found in~\cite{tellez} for a system with
$N'=N-\ell-r=N-m-M$ counterions confined between charges $Q_1=\frac{N}{2}-\ell$ and
$Q_2=\frac{N}{2}-r$. We have defined $Q_> = \max(Q_1,Q_2)=\frac{N}{2}
- m$ and $Q_< = \min(Q_1,Q_2)=\frac{N}{2} - M$. Note that $Q_1$ and
$Q_2$ are the global charges of the colloidal particles plus the
counterions outside the corresponding edge. The relation between the two ensembles
comes from the nature of the one dimensional Coulomb electric
field. As far as the electric field is concerned, the only thing that matters is
the net charge at each side of the point where the field is computed,
and not the detailed position of each charge in the system. Then,
it is equivalent to have charges spatially distributed or one
point charge as long as the net charge is the same. The
proportionality factor only adds up a constant factor to the Gibbs
energy, which accounts for the zero pressure (infinite length in the
canonical ensemble) energy of the additional unbounded counter ions.

\subsection{Gibbs Free Energy and Optimal Configuration}

\subsubsection{General results}
We now turn our attention to the Gibbs energy which will allow us to
determine the fundamental configuration (minimum Gibbs energy
configuration). First we consider the even case $N=2n$. The Gibbs
energy is given by the usual expression $ \widetilde{G} = - \beta^{-1} \ln
Z_P $. Using the dimensionless free energy $G = \widetilde{G} \beta$, we have
\begin{equation}\label{evengibb}
G_{2n}(M,m) =\begin{dcases}
   2 \ln(M!m!) +  \ln P +  2\sum_{k=1}^{n-M}  \ln({P+k^2}) + \sum_{k=n+1-M}^{n-m}\ln({P+k^2} )     & \text{if \;}  n-M-\frac{1}{2} > 0\\
   2 \ln(M!m!) + \sum_{k=M-n}^{n-m}  \ln({P+k^2}) &  \text{if \;} \frac{1}{2}+M-n>0
\,.
\end{dcases}
\end{equation}

First we examine the situation when the total number of unbounded ions is fixed. To this end, consider the
exchange of one particle from one side to the other
\begin{equation}
\label{fixedincrease} \Delta G_{2n}(M \rightarrow M+1, m\rightarrow m-1) =   2\ln\left(\frac{M+1}{m}\right) + \ln \left(\frac{P + (n+1-m)^2}{P+(n-M)^2}\right)
\,,
\end{equation}
\begin{equation}
\label{fixedecrease} \Delta G_{2n}(M \rightarrow M-1, m\rightarrow m+1) =   2\ln\left(\frac{m+1}{M}\right) + \ln \left(\frac{P + (n+1-M)^2}{P+(n-m)^2}\right)
\,.
\end{equation}
As $ n + 1 - m > n - M $ and $n + 1 - M \leq n - m $, from
(\ref{fixedincrease}) and (\ref{fixedecrease}) it can be concluded
that $\Delta G_{2n}(M \rightarrow M+1, m\rightarrow m-1) > 0$ and $
\Delta G_{2n}(M \rightarrow M-1, m\rightarrow m+1) \leq 0$. This means
that the configuration where there are the same amount of left and
right charges $m=M$ is the one that minimizes the Gibbs energy. The
most probable configuration when the number of
unbounded particles is fixed is for $m=M$, or in other words,
$r=l$.

Now we consider transitions where an extra particle becomes unbounded,
that is $M\rightarrow M+1$ at fixed $m$ or $m\rightarrow m+1$ at fixed $M$. The Gibbs free energy
differences are
\begin{equation}
\label{Mincrease} \Delta G_{2n}(M \rightarrow M+1, m\rightarrow m) =  \ln \left(\frac{(M+1)^2}{P+(n-M)^2}\right)
\,,
\end{equation}
\begin{equation}
\label{msinc} \Delta G_{2n}(M \rightarrow M, m\rightarrow m+1) =  \ln \left(\frac{(m+1)^2}{P+(n-m)^2}\right)
\,.
\end{equation}
From these two last expressions, it appears that for large enough
value of $P$, the Gibbs energy difference will always be negative
regardless of the values of $n,m \text{ and } M$. This means that for
a regime of high pressures the configuration where all particles are
between the colloids is the one with highest energy and thus the least
probable. This can be seen physically as follows: high pressures
imply small volume (length in this case). If the ions are confined
tight together, the entropic cost of confinement becomes overwhelming,
and it is more favorable to have ions unbounded, in the leftmost or rightmost regions.
This can be seen as a phenomenon of counterion evaporation.

For small pressures we analyze first the equality
(\ref{Mincrease}). The following inequality guarantees an endergonic
($\Delta G\geq0$) reaction
\begin{equation}
\label{ineq-pre0}   M \geq  \frac{P+n^2-1}{2(n+1)}
\,.
\end{equation}
On the other hand, $2n-m-M$ need to be a positive integer since it is
the number of particles in the inner region between the colloids. Therefore $2n-1-m\geq M$. Putting this together with (\ref{ineq-pre0}), we have $\Delta G_{2n}(M\rightarrow M+1,m\rightarrow m)\geq 0$ when
\begin{equation}
\label{ineq}
 2n-1-m \geq   M \geq  \frac{P+n^2-1}{2(n+1)}
\,.
\end{equation}
From this relation, we can define a special value of the pressure
\begin{equation}
  \label{eq:PH-def}
  P_H(m) =3n^2 + 2n(1-m) - (2m+1)=\frac{3}{4}N^2+N(1-m)-(2m+1)
  \,,
\end{equation}
which satisfies $2n-1-m = \frac{P_H(m)+n^2-1}{2(n+1)}$ and will be
important in the following analysis. Suppose now that $M$ is small
enough such that (\ref{ineq}) is not satisfied, and therefore by
increasing it, while the other parameters $P$ and $m$ are kept fixed,
the Gibbs energy will decrease. One can continue to take out ions
successively from the inner region to the outer region with the
largest number of ions ($M\rightarrow M+1$) and decrease the Gibbs
energy until $M$ reaches a value that satisfies (\ref{ineq}). From
there, increasing $M$ will start to increase the Gibbs
energy. Therefore the value of $M=M_{2n}^c(P,m)$ for which the system
reaches the minimum Gibbs energy, at given pressure $P$ and value of
$m$, is
\begin{equation}
\label{mestre}   M_{2n}^c(P,m) =
\begin{cases}
\ceil*{\frac{P+n^2-1}{2(n+1)}} & \text{if\ }P\leq P_H(m)\\
2n-m & \text{if\ } P>P_H(m)
\text{\ (no ions remain in the inner region),}
\end{cases}
\end{equation}
where $\ceil*{x}$ is the ceiling function.

From relation (\ref{msinc}) one can obtain more information in a
similar fashion. When $m\rightarrow m+1$, $\Delta G\geq 0$ if
\begin{equation}
\label{ineq2}  n-1 \geq m \geq  \frac{P+n^2-1}{2(n+1)}
\,.
\end{equation}
The first inequality stems from the fact that $2n-M-m\geq 1$ and $M\geq m$.
Now an intermediate regime is obtained from this last
inequality. Let us define $P_I$ such that $\frac{P_I+n^2-1}{2(n+1)}=n-1$, that is
\begin{equation}
  \label{eq:PI-def}
  P_I=n^2-1
  \,.
\end{equation}
If $P_I< P \leq P_H$ and $M$ satisfies (\ref{ineq}), then freeing
ions to the side which contains largest amount of ions ($M\rightarrow
M+1$ and $m\rightarrow m$) will increase the Gibbs energy. However if
ions are freed to the other side (the one with the smallest amount of
ions, $m\rightarrow m+1$ and $M\rightarrow M$), the Gibbs energy will
decrease.

\subsubsection{Optimal ionic configurations	}
\label{sec:optimal}
We are now in a position to identify the fundamental energy/configuration. Note
that we only need to compare the energies of the configurations which
have $M=m$ or $M=m+1$. These were found to be the
configurations with minimal energy among the systems with a fixed number of free
ions. First let us compare the transitions in which the number of free
ions is kept even, that is $M\rightarrow M+1$ and $m\rightarrow m+1$
with $M=m$. This is given by
\begin{equation}
\label{prooffund} \Delta G_{2n}(M \rightarrow M+1, m\rightarrow m +1) =  2\ln \left(\frac{(m+1)^2}{P+(n-m)^2}\right)\,.
\end{equation}
The argument of the logarithm is the same as in (\ref{Mincrease}). Then following the same analysis done for equation (\ref{Mincrease}) but restricting the value of $m$ such that $m<n-1$, we find that the value for which $m$ minimizes the energy is
\begin{equation}
\label{mint}   m_{2n}^c(P) =
\begin{cases}
\ceil*{\frac{P+n^2-1}{2(n+1)}} & \text{if\ } P\leq P_I \\
n  & \text{if\ } P>P_I\,.
\end{cases}
\end{equation}
Since $m_{2n}^c(P)$ and $M_{2n}^c(P)$ are the same when $M=m$, it
follows that any configuration with an odd number of unbounded
particles must be more energetic than the configuration with
$M=m=m_{2n}^c(P)$. Then the fundamental configuration has an even
number of unbounded ions, with $M=m=m_{2n}^c(P)$.

\begin{figure}[h]
  \centering
  \includegraphics[width=.9\linewidth]{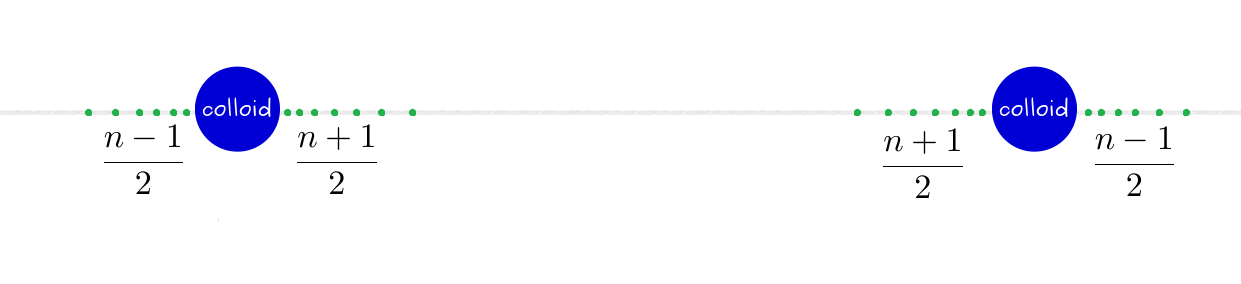}
  \caption{Fundamental configuration for $P=0$
    and $N=2n=26$ (even), $n=13$ (odd). In this case, the
    first threshold is precisely $P_0=0$. As soon as $P$
    increases and $0<P\leq P_1$, one particle from each
    interior side will go to the outside, leaving
    $(n-1)/2$ particles in each interior side of the
    colloid ans $(n+1)/2$ on each outer side.}
  \label{fig:N26}
\end{figure}

\begin{figure}
  \centering
  \includegraphics[width=0.9\linewidth]{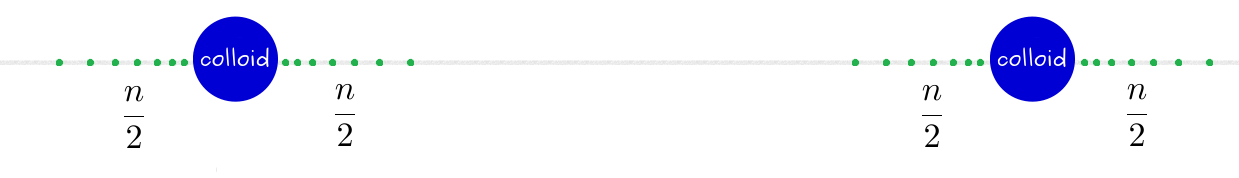}
  \caption{Fundamental configuration for $P=0$
    and $N=2n=28$ (even), $n=14$ (even).}
  \label{fig:N28}
\end{figure}

The evolution of the fundamental configuration as $P$ varies is the
following. For $P=0$, which corresponds to a large average separation
$\langle L \rangle$ between the colloids, the number of unbound ions
on both sides of the colloids is
\begin{equation}
  m_{2n}^c(0)=\ceil*{\frac{n-1}{2}}=
  \begin{cases}
    n/2 & \text{if $n$ is even,}\\
    (n-1)/2 & \text{if $n$ is odd.}
  \end{cases}
\end{equation}
The remaining bound ions will be divided in two and locate themselves
in the vicinity of each colloid. This number of bound ions around each
colloid is
\begin{equation}
  n-m_{2n}^c(0)=
  \begin{cases}
    n/2 & \text{if $n$ is even,}\\
    (n+1)/2 & \text{if $n$ is odd.}
  \end{cases}
\end{equation}
These configurations are shown in Figs.~\ref{fig:N26}
and~\ref{fig:N28}. Essentially, there is one quarter of the total
number of counterions on each side of each colloid. For the case when
$n$ is even it is exactly one quarter ($n/2=N/4$) and in the case when
$n$ is odd there is a frustration to achieve this and there is one
counterion more on each inside side than on the outside sides. This
configuration can be understood by a simple argument. For $P=0$,
$\langle L \rangle \to \infty$, so each colloid is like an isolated
system that will attract $N/2$ counterions to neutralize it. Since the
effect of the other colloid will be negligible, the left and right
sides of each colloids are equivalent and the $N/2=n$ counterions will
distribute themselves in equal amounts around each side (parity of $n$
permitting).


Then as $P$ increases, $\langle L \rangle$ decreases, and it becomes
more entropically favorable for the counterions in the inside region
to ``jump out'' to the outside regions~\footnote{We use the word
  ``jump'' figuratively, since the system is really one-dimensional,
  and the particles go thru the colloids at each transition rather
  than jumping over them.}.
There is no symmetry breaking
between the left and right outside regions, so at each transition two
counterions simultaneously jump to the outside, one on each
side. These transitions occur when the argument of the ceiling
function in Eq.~(\ref{mint}), $(P+n^2-1)/(2(n+1))$, is an integer. The
values of the pressure at which a transition occurs ($P=P_k$) can be
indexed by an integer $k$. Table~\ref{tab:transitions} shows the
values of $P_k$ which depend on the parity of $N$ and $n$. At each
transition $P=P_k$ there is a 4-fold degeneracy where the configurations
$(\ell,r)$ corresponding to $(m_{2n}^c(P_k)-1, m_{2n}^c(P_k)-1)$,
$(m_{2n}^c(P_k), m_{2n}^c(P_k)-1)$, $(m_{2n}^c(P_k)-1, m_{2n}^c(P_k))$,
$(m_{2n}^c(P_k), m_{2n}^c(P_k))$ have all the same Gibbs energy.

\begin{table}
  \caption{Values of the pressure $P_k$ at which a jumping transition
    occurs.}
  \label{tab:transitions}
  \begin{tabular}{|c||c|c||c|c|}
    \hline
    & \multicolumn{2}{|c||}{$N=2n$} & \multicolumn{2}{c|}{$N=2n+1$}\\
    \hline
    & $n$ even & $n$ odd & $n$ even & $n$ odd \\
    \hline
    Pressure threshold $P_k$ &
    \ $(N+2)(k+\frac{1}{2})$ \  & $(N+2) k$ &
    \ $(N+2)(k+\frac{1}{4})$ \ &  $(N+2)(k+\frac{3}{4})$ \\
    \hline
    \ Number of unbound ions $M_{N}^c$ \ &
    $\frac{n}{2}+k$ & $\frac{n-1}{2}+k$ &
    $\frac{n}{2}+k$ & $\frac{n+1}{2}+k$ \\
    \hline
    Index $k$ range &
    \ $k\in\{0, 1,\ldots, \frac{n}{2}-1\}$ \  &
    \ $k\in\{0, 1,\ldots, \frac{n-1}{2}\}$ \ &
    \ $k\in\{0, 1,\ldots, \frac{n}{2}\}$ \ &
    \ $k\in\{0, 1,\ldots, \frac{n-1}{2}\}$ \ \\
    \hline
  \end{tabular}

\end{table}

These transitions continue as $P$ increases until all particles are
outside. This occurs for $P=P_I=n^2-1$. At this value, the last two
counterions that are in the inside region jump to the outside
region. The corresponding Gibbs energy of these configurations are
\begin{equation}
  G_{2n}(n-1,n-1)=4\ln (n-1)! + 2 \ln (P+1) + \ln P
  \qquad \text{(there remains only two counterions inside),}
\end{equation}
and
\begin{equation}
  G_{2n}(n,n)=4\ln n! + \ln P
  \qquad
  \text{(all counterions out).}
\end{equation}
When $P=P_I=n^2-1$ we have $G_{2n}(n-1,n-1)=G_{2n}(n,n)=G_{2n}(n,n-1)$.

\begin{figure}[h]
	\centering
	\begin{minipage}{.5\textwidth}
		\centering
		\includegraphics[width=0.9\linewidth]{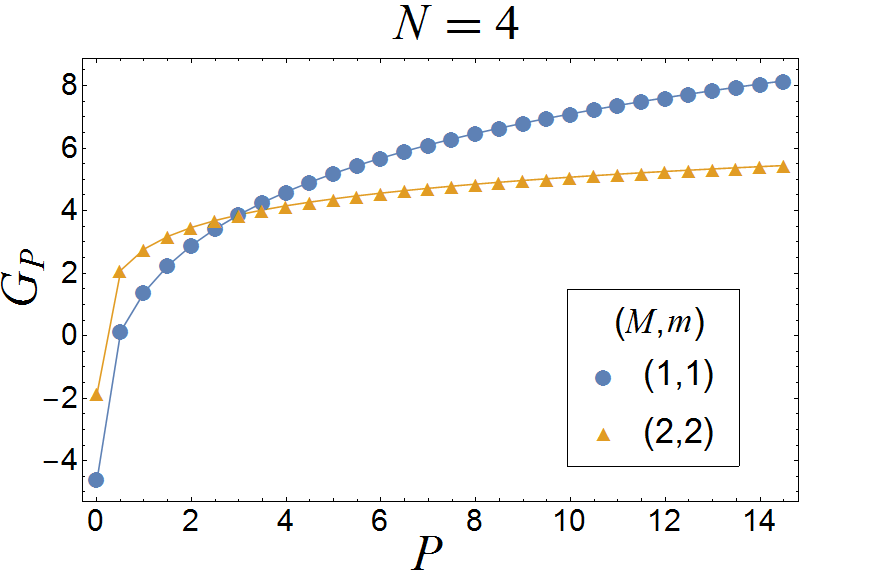}
		\captionof{figure}{Gibbs energies for $N=4$. The solid
                  lines represent the configurations with minimal
                  Gibbs energy for a fixed number of unbounded
                  particles.}
		\label{gpN4}
	\end{minipage}%
		\begin{minipage}{.5\textwidth}
			\centering
			\includegraphics[width=1.0\linewidth]{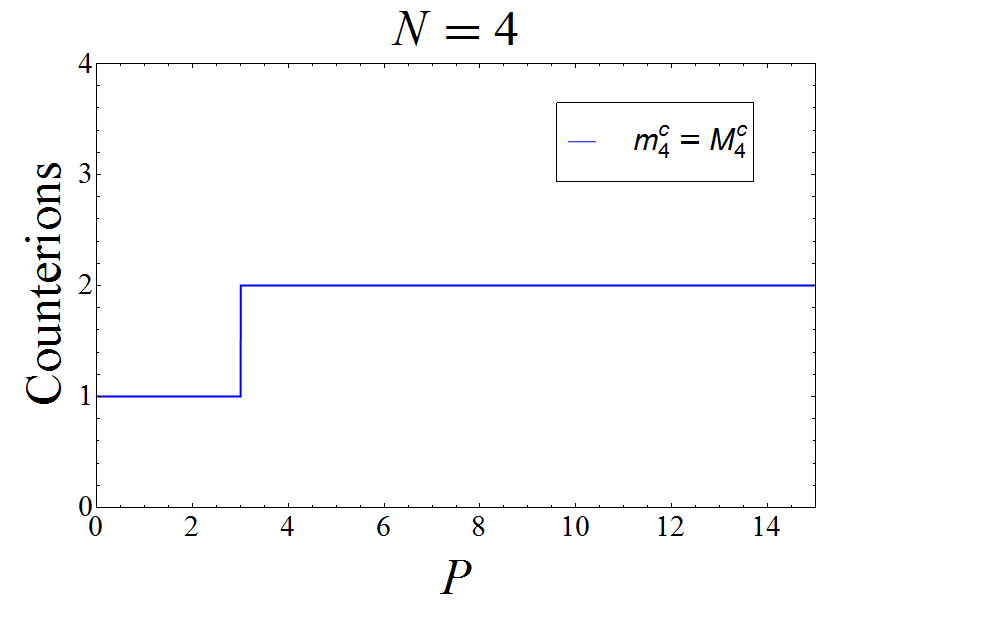}
			\captionof{figure}{Configuration that
                          minimizes the Gibbs energy for $N=4$. The
                          figure shows
			the value of $m=M$, which is half the number of unbounded counter-ions.}
			\label{mpN4}
		\end{minipage}
\end{figure}

\begin{figure}[h]
	\centering
	\begin{minipage}{.5\textwidth}
		\centering
		\includegraphics[width=1.05\linewidth]{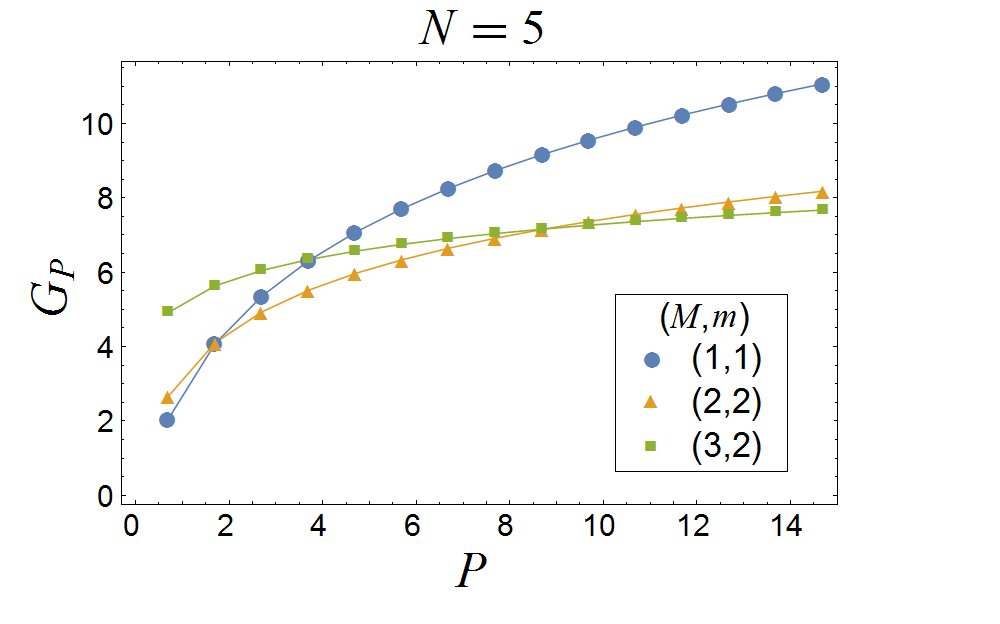}
		\captionof{figure}{ Gibbs energies $N=5$. The solid
                  lines represent the configurations with minimal
                  Gibbs energy for a fixed number of unbounded
                  particles.}
		\label{gpN5}
	\end{minipage}%
	\begin{minipage}{.5\textwidth}
		\centering
		\includegraphics[width=1.05\linewidth]{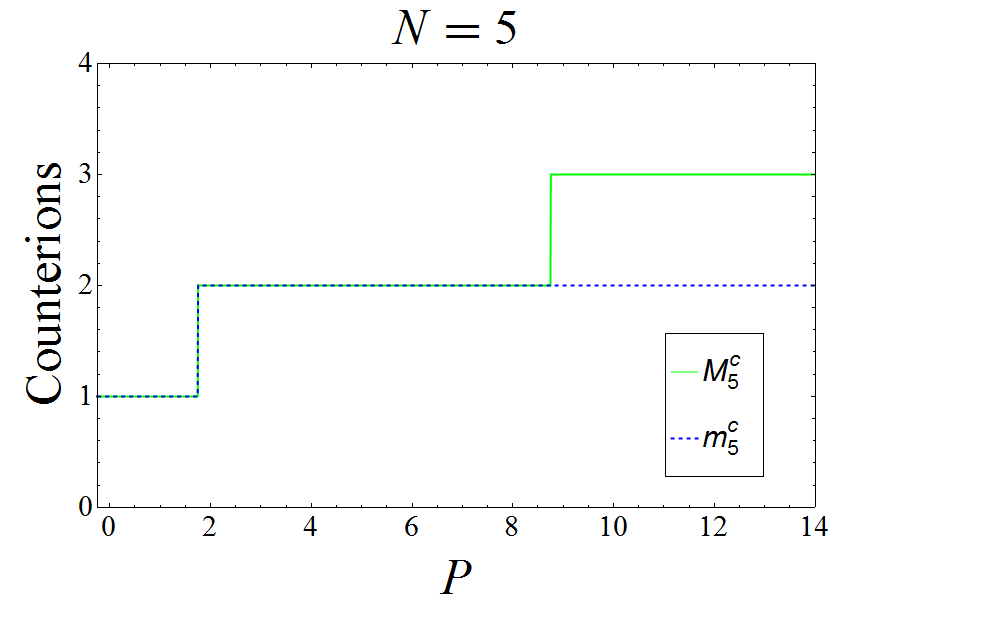}
		\captionof{figure}{Configuration that minimizes the Gibbs energy $N=5$. The values of both $m$
		and $M$ are reported ; the total number of unbounded ions is $m+M$.}
		\label{mpN5}
	\end{minipage}
\end{figure}

Figs.~\ref{gpN4}
and \ref{mpN4} illustrate this situation when $N=4$ ($n=2$). In that
case $P_I=3$, which is the value of the pressure above which it is more
favorable to take all four particles unbounded outside (two on each side), than
to have two outside and two confined in the inner region.
The discussion of the present results comes with a word of caution. When
$n=M=m$, as for the curve labeled (2,2) on Fig. \ref{gpN4}, the force
felt by each of the ``colloids'' at $x=0$ and $x=L$ vanishes. This is because
the electric field acting on a colloid (say at $x=0$), reads simply $-n$
(it tends to repel the colloid from the other located at $x=L$),
so that the resulting force is $-n^2$.
On the other hand,  the osmotic pressure stemming from counterions on the left hand side
of the colloid,
creates a contribution $n^2$, that is exactly opposite. The fact that the pressure vanishes
makes that the isobaric ensemble becomes more subtle to analyze, and that the quantity
$P$ involved in the Laplace transformation to obtain the partition function
is not the physical pressure of the system. As a consequence,
the curve $(2,2)$ on Fig. \ref{gpN4}, which admits a simple analytical expression
($G_P=\ln(16\,P)$), indicates by its domain of prevalence ($P>3$), a region that is
physically forbidden (in a canonical description). We shall see when discussing canonical ensemble results,
that this scenario is confirmed, and that the pressure is indeed always smaller than 3
in the case $N=4$. A similar phenomenon appears when $N$ is odd, and all ions are
unbounded (meaning that $N=2n+1$, $M=n+1$ and $m=n$). In this case,
one can show that $P$ is again $L$ independent (and more precisely, that
$P=-1/4$, irrespective of $N$); the canonical and isobaric descriptions yield distinct results:
\emma{when the canonical pressure is size independent, the system cannot adjust
its volume to adapt to the externally imposed pressure $P$. The isothermal-isobaric ensemble
then exhibits an instability, leading the system to adopt a vanishing volume, or an infinite
one, dependent on $P$. This suppresses the equivalence of ensembles.
One can be slightly more precise and state that the ensemble equivalence is lost when the canonical free energy
does not diverge for $L=0$.}
We therefore anticipate that for the $N=5$ results to be shown below (Fig.
\ref{gpN5}), the domain of prevalence of the (3,2) configuration also signals
a region that is, canonically, unphysical.


%
%
%


Now, the odd case ($N=2n+1$) is examined. The dimensionless Gibbs energy is given by the following expression:
\begin{equation}\label{oddgibb}
G_{2n+1}(M,m) = \begin{dcases}
  2 \ln(M!m!) +   2\sum_{k=0}^{n-M}  \ln\left({P+\left(k+\frac{1}{2}\right)^2} \right) + \sum_{k=n+1-M}^{n-m}\ln\left({P+\left(k+\frac{1}{2}\right)^2} \right)      & \text{if \;\;}  n-M+\frac{1}{2} > 0 \\
 2 \ln(M!m!) +   \sum_{k=M-n-1}^{n-m} \ln\left({P+\left(k+\frac{1}{2}\right)^2} \right) &  \text{if\;\;}  M-\frac{1}{2}-n>0
\end{dcases}
\end{equation}
%
The energy changes for a fixed number of free ions are given by the following:
\begin{equation}
\label{oddfix1}  \Delta G_{2n+1}(M\rightarrow M+1 , m \rightarrow m-1) = 2 \ln \left(\frac{M+1}{m}\right) + \ln \left(\frac{P + (n-m +3/2)^2}{P + (n-M+1/2)^2}\right) \geq 0
\,,
\end{equation}
\begin{equation}
\label{oddfix2}  \Delta G_{2n+1}(M\rightarrow M-1 , m \rightarrow m+1) = 2 \ln \left(\frac{m+1}{M}\right) + \ln \left(\frac{P + (n-M +3/2)^2}{P + (n-m+1/2)^2}\right) \leq 0
\,.
\end{equation}
From these expressions we obtain again a behavior in which the system
prefers to be with the same amount of counter ions on each side. In
case the number of free ions is odd, the fundamental configuration is
degenerated because changing the exceeding ($m=M-1$) ion from left to right
is indifferent to the Gibbs energy.

The Gibbs energy differences when only $M$ or $m$ is changed are
\begin{equation}
\label{oddunfix1}  \Delta G_{2n+1}(M\rightarrow M+1 , m \rightarrow m) =\ln \left(\frac{(M+1)^2}{P + (n-M+1/2)^2}\right)
\,,
\end{equation}
\begin{equation}
\label{oddunfix2}  \Delta G_{2n+1}(M\rightarrow M , m \rightarrow m+1) =\ln \left(\frac{(m+1)^2}{P + (n-m+1/2)^2}\right)
\,.
\end{equation}
Performing a similar analysis as in the even case we conclude that $M\rightarrow M+1$ with $m$ fixed yields $\Delta G\geq 0$ if
\begin{equation}
  2n-m\geq M \geq \frac{P+n^2+n-3/4}{2n+3}
  \,.
\end{equation}
We can define a special value of the pressure as before: $P_H(m) = 3
n^2 + 5n + 3/4 -m(3+2n)$. In terms of $N=2n+1$, this is the same
as~(\ref{eq:PH-def}). The $M$ such that the system has the minimum
energy, $M_{2n+1}^c(P)$, for a given pressure and a given $m$ is
\begin{equation}
  \label{mestreodd}
  M_{2n+1}^c(P) =
  \begin{cases}
    \ceil*{\frac{P+n^2 +n-3/4}{3+2n}} & \text{if\ }  P\leq P_H\,,\\
     2n+1 - m & \text{if\ }P>P_H
    \,.
  \end{cases}
\end{equation}

On the other hand, when $m\rightarrow m+1$ with $M$ fixed, we have $\Delta G\geq 0$ if
\begin{equation}
  n \geq m \geq \frac{P+n^2+n-3/4}{2n+3}
  \,.
\end{equation}
Let us define $P_I$ such that $\frac{P_I+n^2+n-3/4}{2n+3}=n$, that is
\begin{equation}
  P_I=n^2+2n +3/4
  \,.
\end{equation}
The same analysis done to obtain the fundamental configuration for
$N=2n$ is valid for the case $N=2n+1$. The fundamental
configuration for $P\leq P_I$ is given by $M=m=
m_{2n+1}^c(P)$ with
\begin{equation}
  \label{mestreodd2}
  m_{2n+1}^c(P) =
  \begin{cases}
    \ceil*{\frac{P+n^2 +n-3/4}{3+2n}} & \text{if\ } P\leq P_I
    \,, \\
    n & \text{if\ }
    P>P_I
    \,.
  \end{cases}
\end{equation}

\begin{figure}[h]
  \centering
  \includegraphics[width=0.9\linewidth]{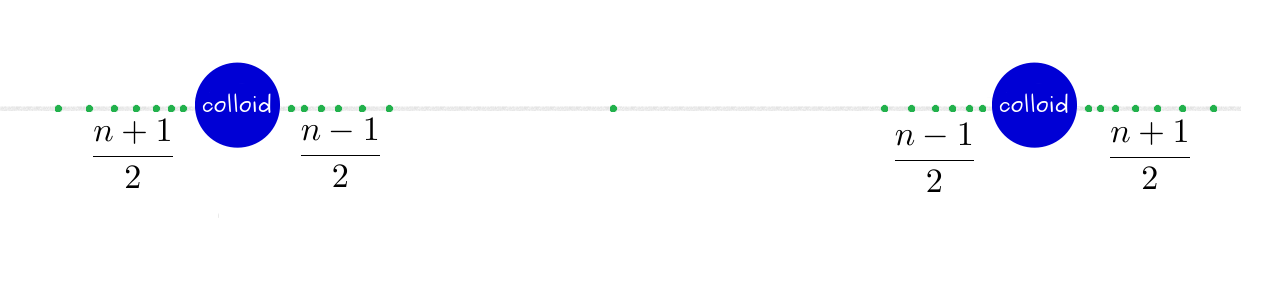}
  \caption{Fundamental configuration for $P=-1/4$ \emma{(meaning that inter-colloid distance is large)}
    and $N=2n+1=27$ (odd), $n=13$ (odd).}
  \label{fig:N27}
\end{figure}

\begin{figure}
  \centering
  \includegraphics[width=0.9\linewidth]{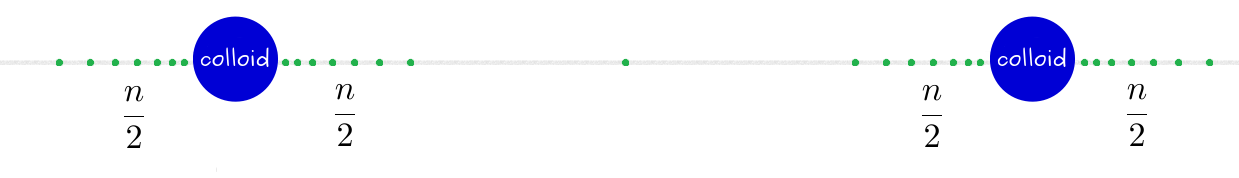}
  \caption{\emma{Same as Fig. \ref{fig:N27}, with still an odd $N=2n+1$, but now $n$ even:} $P=-1/4$,
    $N=29$, and $n=14$.}
  \label{fig:N29}
\end{figure}

The evolution of the fundamental configuration as $P$ increases is
similar to the case $N$ even. The smallest physical value for the
pressure is $P=-1/4$ which corresponds to $\langle L \rangle \to
\infty$. As before the counterions will arrange themselves in four
quarters around each side of the colloids. Nevertheless, since
$N=2n+1$ there is a ``misfit'' counterion that roams in the inside
region between the two colloids, responsible for the effective
attractive force ($P=-1/4$) between the colloids. The role of this
counterion is analyzed in more detail in
Ref.~\cite{tellez}. Figs.~\ref{fig:N27} and~\ref{fig:N29} show the
possible fundamental configurations for $P=-1/4$.


As $P$ increases, the transitions where two inside counterions jump to
the outside region will occur each time ${(P+n^2 +n-3/4)}/{(3+2n)}$ is
an integer, accompanied by the 4-fold degeneracy previously
discussed. Notice that the first transition will occur for a value of
$P>0$, therefore the configuration discussed in the previous paragraph
is the fundamental one for all the region of attractive effective
force ($P<0$) and also for small
pressures. Table~\ref{tab:transitions} shows the values of the
pressure at the transitions thresholds and the number of counterions
$M_{2n+1}^c$ outside.

Finally when $P=P_I=n^2+2n+3/4$ we have
$G_{2n+1}(n,n)=G_{2n+1}(n+1,n)$. The last inside counterion will jump
to one of the outside regions.  For $P>P_I$ the lowest energy is
degenerated for the configurations determined by $M=n+1$ and $m=n$.

The situation is illustrated in Figs.~\ref{gpN5} and \ref{mpN5} for $N=5$ ($n=2$). In this case $P_I=35/4=8.75$. Equation (\ref{mestreodd2}) predicts two transitions: first from having one particle at each outer side to having two particles at each outer side (at $P_0=1.75$), then the remaining particle goes to one outer region when $P=P_1=P_I=8.75$.

\subsection{Isobaric Length }

The isobaric length is given by the usual relation $\braket{L} =- \frac{\partial \ln Z_p }{\partial P} $:

\begin{equation}\label{Lpar}
\braket{L}_{2n}(M,m) = \begin{dcases}
\frac{1}{P} +  2\sum_{k=1}^{n-M}  \frac{1}{P+k^2}  + \sum_{k=n+1-M}^{n-m}\frac{1}{P+k^2}  & \text{if\;\;} n - M -\frac{1}{2} > 0\\
\sum_{k=M-n}^{n-m} \frac{1}{P+k^2}  & \text{if\;\;} \frac{1}{2} + M - n > 0
\end{dcases}
\end{equation}

%

\begin{equation}\label{Lodd}
\braket{L}_{2n+1}(M,m) = \begin{dcases}
  \sum_{k=n+1-M}^{n-m} \frac{1}{{P+\left(k+\frac{1}{2}\right)^2} } + 2\sum_{k=0}^{n-M}   \frac{1}{{P+\left(k+\frac{1}{2}\right)^2} } & \text{if\;\;}   n - M +\frac{1}{2}>0 \\
 \sum_{k=M-n-1}^{n-m} \frac{1}{{P+\left(k+\frac{1}{2}\right)^2} } & \text{if \;\;} M -\frac{1}{2}- n>0
\end{dcases}
\end{equation}



\begin{figure}[h]
	\includegraphics[width=1\linewidth]{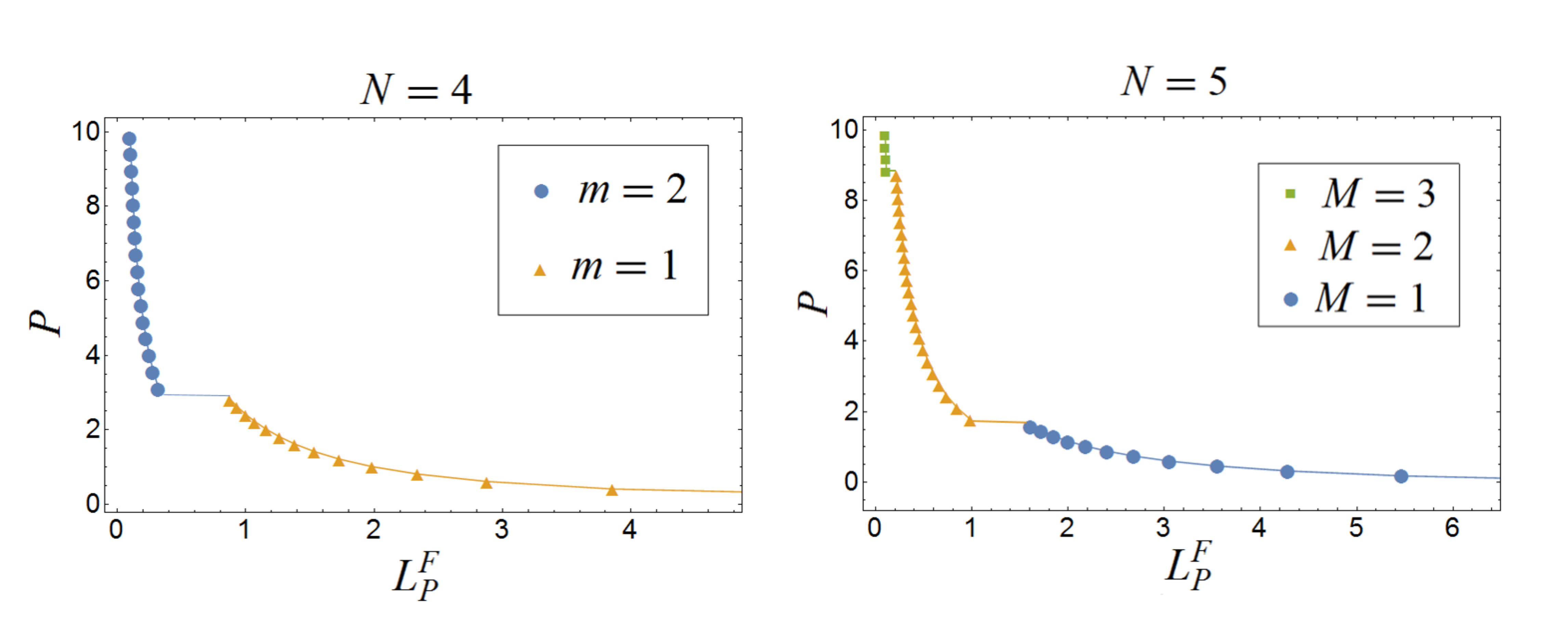}
	\caption{Fundamental isobaric length $L_P^F = \langle L \rangle$ for $N=4$ (left) and $N=5$ (right) }
		\label{plf}
\end{figure}

We define $L_P^{F}$ as the value of $\langle L \rangle$ corresponding
to the minimal Gibbs energy configuration for a given pressure
$P$. This quantity $L_P^{F}$ is obtained from Eqs.~(\ref{Lpar})
and~(\ref{Lodd}) by replacing the appropriate values of $m$ and $M$
that correspond to the minimal Gibbs energy configuration.  By doing
this, we are considering a situation where the system is quenched at
that fundamental configuration. In Fig.~\ref{plf} we show
graphically the relation between the length $L_P^F$ and the pressure
$P$.

\begin{figure}[h]
	\includegraphics[width=1\linewidth]{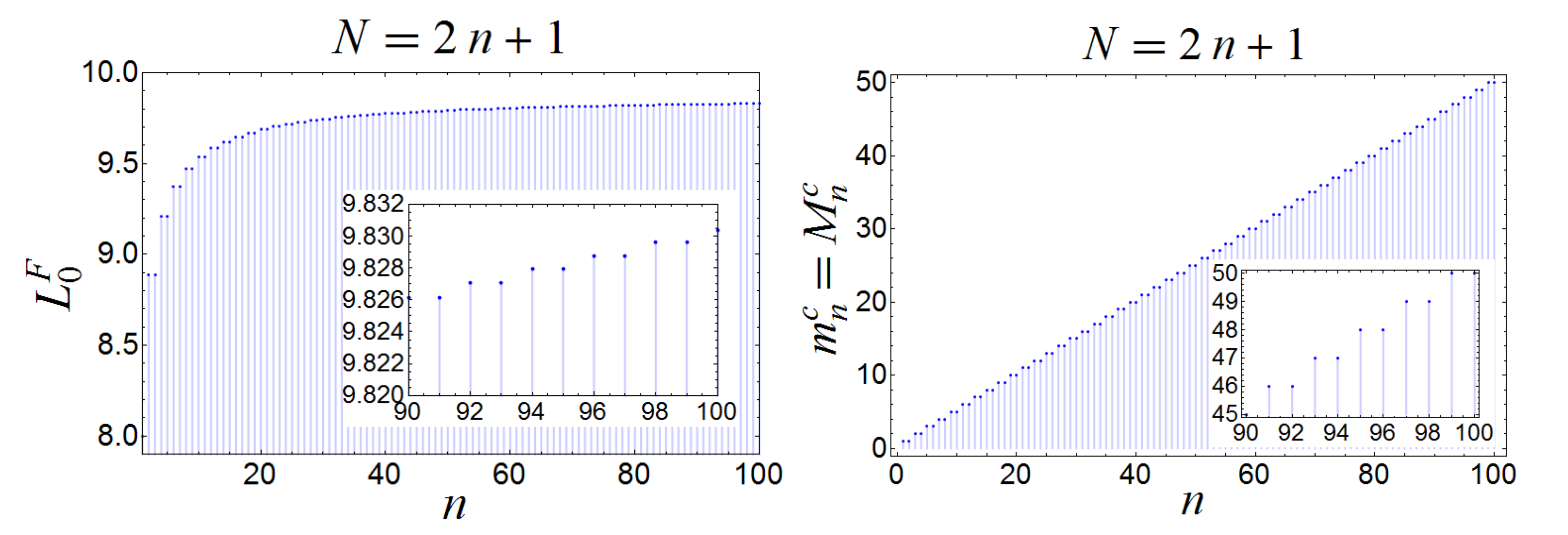}
	\caption{Left: Intersection of $L_p^F(n)$ with $P=0$ as a function of $n$. $L_p^F(n) \in[8,10)$,
	approaching asymptotically 10. Right: Corresponding values of $m_n^c = M_n^c$ as a function of $n$.
	For both graphs, the insets are zooms, emphasizing the large $n$ region, where a doublet
	structure is appararent. }
	\label{plff}
\end{figure}

The behavior of the fundamental isobaric length shown in Fig.\ref{plf} is consistent with the physical intuition for small pressures. The only unexpected behavior is the
asymptotic approach to zero length as $P$ tends to infinity. It would be expected that it reaches zero (for even $N$), or $-1/4$ (for $N$ odd) for a finite value of $P$,
when all particles become unbounded, as discussed after Fig. \ref{gpN4}.
The fact that this does not happen is due to the characteristic that the pressure of the isobaric ensemble includes the right colloidal charge, unlike the canonical ensemble.
We assume here, for the sake of the argument, that the left colloid is held fixed without fluctuations allowed.
Thus, a barostat acting on the right-most colloid has to work against the fluctuations of the colloid itself,
which contributes to the barostatic pressure.
\emma{As alluded to above, we emphasize that the branch with $P>3$
in Fig. \ref{plf}, having all counterions out, is such that the
equivalence of canonical and isobaric ensembles is lost. This will be confirmed
in Fig. \ref{fp} where it will appear that the value of 3 is the upper bound
for the canonical pressure.}
Note also that for $N=4$ the pressure never reaches negative values,
while for $N=5$ it is negative for $L_p^F>80/9$. For an arbitrary $n$
such that $N=2n+1$ we have a similar behavior with $P$ becoming
negative for a $L_0^F(n)$.  The function $L_0^F(n)$ as a function of
$n$ is shown in Fig.~\ref{plff} and it can be obtained exactly by
replacing $P=0$ in equation (\ref{Lodd}). It is a monotonically
increasing function for $n\in \mathbb{N}^*$, bounded from below and
from above by 8 and 10 respectively. The lower bound is realized for
$n=1$. Note that each value of $L_0^F$ is repeated twice when $n$ is
increased by one, a behavior that can be explained by analyzing the
corresponding number of unbound particles.  When $n$ is increased to
$n+1$ and the number of unbounded particles remains the same $m^c_{n}
= m^c_{n+1}$, in the system of $n+1$ there are two additional
particles in the bounded region. The only way to maintain the same
pressure with more bounded particles is by increasing the length. Then,
for a $n$ such that $m^c_{n} = m^c_{n+1}$ the length for $n+1$
increases $ L_0^F(n)<L_0^F(n+1)$. If on the other hand the number of
unbounded particles increases, there are the same amount of
counterions inside leaving the pressure and length unchanged. Remember
that the fundamental configurations are for $M = m$ and thus if $m$
increases by 1, so does $M$. This means that either both particles
become unbounded or none at all are. For $N=2n$, the fundamental
pressure is always positive. Consequently, like-charge attraction only
occurs for an odd number of counterions.

%

\section{Canonical Ensemble }\label{III}

\subsection{Canonical partition function}

The canonical partition function is obtained by computing the inverse Laplace transform of equation (\ref{isobaric}). This is done with the aid of the residue theorem.
It is more convenient to work with expressions (\ref{barieven}) and (\ref{bariodd}), from which it is easier to identify the first and second order poles of the isobaric partition function.
For $M < n$ and $N=2n$ (even number of counter ions) there are both first and second order poles, which leads to the expression
\begin{multline}\label{zcanonev2polos}
Z_c(2n,L) = \frac{1}{(M!m!)^2} \left[   \frac{1}{[(n-M)!(n-m)!]^2} - \sum_{j= n+1-M}^{n-m}\frac{e^{-j^2L}}{j^2} \left(\prod_{k=1}^{n-M}\frac{1}{j^2-k^2}\right)^2 \prod_{\substack{k=n+1-M\\k \neq j}}^{n-m}\frac{1}{k^2-j^2}   \right. \\
  \left.    -\sum_{j=1}^{n-M}\frac{e^{-j^2L}}{j^2}\left(\prod_{\substack{k=1\\k \neq j}}^{n-M}\frac{1}{j^2-k^2}\right)^2 \left(\prod_{\substack{k=n+1-M}}^{n-m}\frac{1}{k^2-j^2}     \right) \left[ L + \frac{1}{j^2} - 2\sum_{\substack{k=1 \\ k\neq j}}^{n-M} \frac{1}{k^2-j^2} - \sum_{k=n+1-M}^{n-m} \frac{1}{k^2-j^2}  \right]                                               \right]
\,.
\end{multline}
If $M \geq n$ and $N$ is even, the isobaric partition function only contains simple poles, leading to
\begin{equation}\label{zcanonevm}
Z_c(2n,L) =  \frac{1}{(M!m!)^2}\sum_{j= M-n}^{n-m} e^{-j^2 L} \prod_{\substack{k=M-n\\k \neq j}}^{n-m}\frac{1}{k^2-j^2}
\,.
\end{equation}

For the odd case, $N=2n+1$, something analogous happens. For $M \leq n$ there are second order poles and for $M>n$ there are only simple poles as seen in expressions (\ref{zcanonod2polos}) and (\ref{simplepoleodd}) below.
\begin{multline}\label{zcanonod2polos}
Z_c(2n+1,L) = \frac{1}{(M!m!)^2} \left[  \sum_{j= n+1-M}^{n-m}e^{-(j+\frac{1}{2})^2 L} \left(\prod_{k=0}^{n-M}\frac{1}{j^2+j-k^2-k}\right)^2 \prod_{\substack{k=n+1-M\\k \neq j}}^{n-m}\frac{1}{k^2+k-j^2-j}  +  \sum_{j=0}^{n-M}e^{-(j+\frac{1}{2})^2 L} \right. \\
 \times\left.   \left(\prod_{\substack{k=0\\k \neq j}}^{n-M}\frac{1}{k^2+k-j^2-j} \right)^2 \left(\prod_{\substack{k=n+1-M}}^{n-m}\frac{1}{k^2+k-j^2-j}   \right) \left[ L  - \sum_{\substack{k=0 \\ k\neq j}}^{n-M} \frac{2}{k^2+k-j^2-j}  - \sum_{k=n+1-M}^{n-m} \frac{1}{k^2+k-j^2-j}  \right]                                               \right]
\,,
\end{multline}
\begin{equation}\label{simplepoleodd}
Z_c(2n+1,L)  =   \frac{1}{(M!m!)^2}   \sum_{j=M-n-1}^{n-m} e^{-(j+\frac{1}{2})^2L}\prod_{\substack{k = M-n-1\\ k \neq j}}^{n-m}    \frac{1}{(k-j)(k+j+1)}
\,.
\end{equation}

\subsection{Helmholtz Free Energy}\label{freeenergy}

To discuss the configuration that will be adopted in the canonical ensemble, we analyze the Helmholtz free energy, $\widetilde{A}$.
We use the relation $\widetilde{A} = - \beta^{-1} \ln Z_c$ and we introduce the dimensionless free energy $A= \beta \widetilde{A}$. The analytic expression for $A$ is hard to
analyze but its physical interpretation is straightforward. First consider small values of the length $L$. There are two different behaviors as $L$ approaches zero. If all counter ions are unbounded,
they screen the colloids, behaving effectively as two charges of opposite sign decreasing their energy as they get closer.
If at least one particle is bounded, again the two colloids are screened but as the distance between them decreases the pressure increases making $A$ diverge.
Now consider the asymptotic behavior when $L \rightarrow \infty$. If $N$ is odd the energy diverges because the ensemble decouples in two charges of opposite sign.
Separating them requires work, which gives an increase of $A$ as $L$ grows. This due to the fact that $A$ can be interpreted as the energy required to assemble the system.
However if $N$ is even, two behaviors are observed. If $M<n+1$, the
all the ions screen the opposite colloid creating two neutral systems, that will require a finite amount of energy to be separated an infinite distance. If $M \geq n+1$, this screening is not successful and we have the same situation as in case when $N$ is odd. A summary of the behavior of $A$ for $L\to 0$ and $L\to\infty$ is shown in Fig.~\ref{fig:A}.

\tikzstyle{level 1}=[level distance=3.5cm, sibling distance=3.5cm]
\tikzstyle{level 2}=[level distance=3.5cm, sibling distance=2cm]

\tikzstyle{bag} = [text width=5em, text centered]
\tikzstyle{end} = [circle, minimum width=3pt,fill, inner sep=0pt]

\begin{figure}[h]
  \centering
  \begin{tikzpicture}[grow=right,scale=0.84]
\node[bag] {$A$}
child {
	node[bag] { $L\rightarrow 0$ }
	child {
		node[bag] {All ions are unbounded $\ell + r = N$ }
		child {
		node[end, label=right:
		{ $   2\ln(\ell!r!)$}] {}}
	}
	child {
		node[bag] {Some ions are unbounded $\ell + r < N$ }
		child {
			node[end, label=right:
			{ Diverges $\sim (N-\ell-r)\ln \frac{1}{L}$}] {}}
	}
}
child {
	node[bag] { $L\rightarrow \infty$ }
	child {
		node[bag] {$N$ Odd }
		child {
			node[end, label=right:
			{ Diverges $\begin{cases}
			 \sim	L/4&\text{for} \quad M\leq n\\
			\sim	(M-n-1/2)^2 L& \text{for} \quad M> n
				\end{cases}$} ] {}}
	}
	child {
		node[bag] {$N$ Even    }
		child {
			node[bag] {$M \leq n$ }
			child {
				node[end, label=right:
				{  $2\ln(M!m!(n-M)!(n-m)!)$ }] {}}
		}
		child {
			node[bag] {$M > n$ }
			child {
				node[end, label=right:
				{ Diverges $ \sim  (M-n)^2L$}] {}}
		}
	}
};
\end{tikzpicture}
  \caption{Behavior of the Helmoltz free energy.}
  \label{fig:A}
\end{figure}
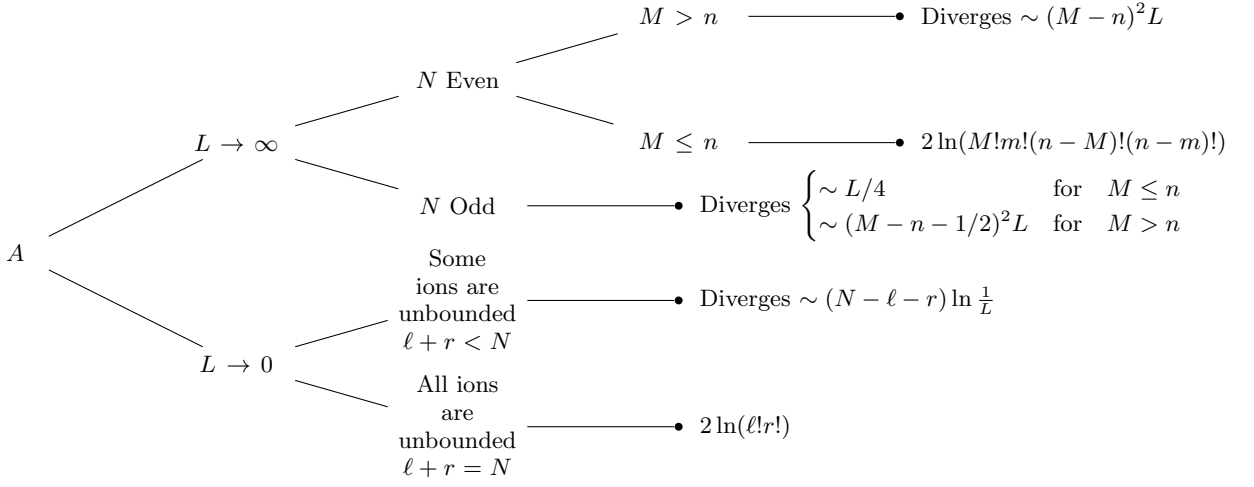

\subsection{Canonical Pressure}

\begin{figure}[h]
	\includegraphics[width=1\linewidth]{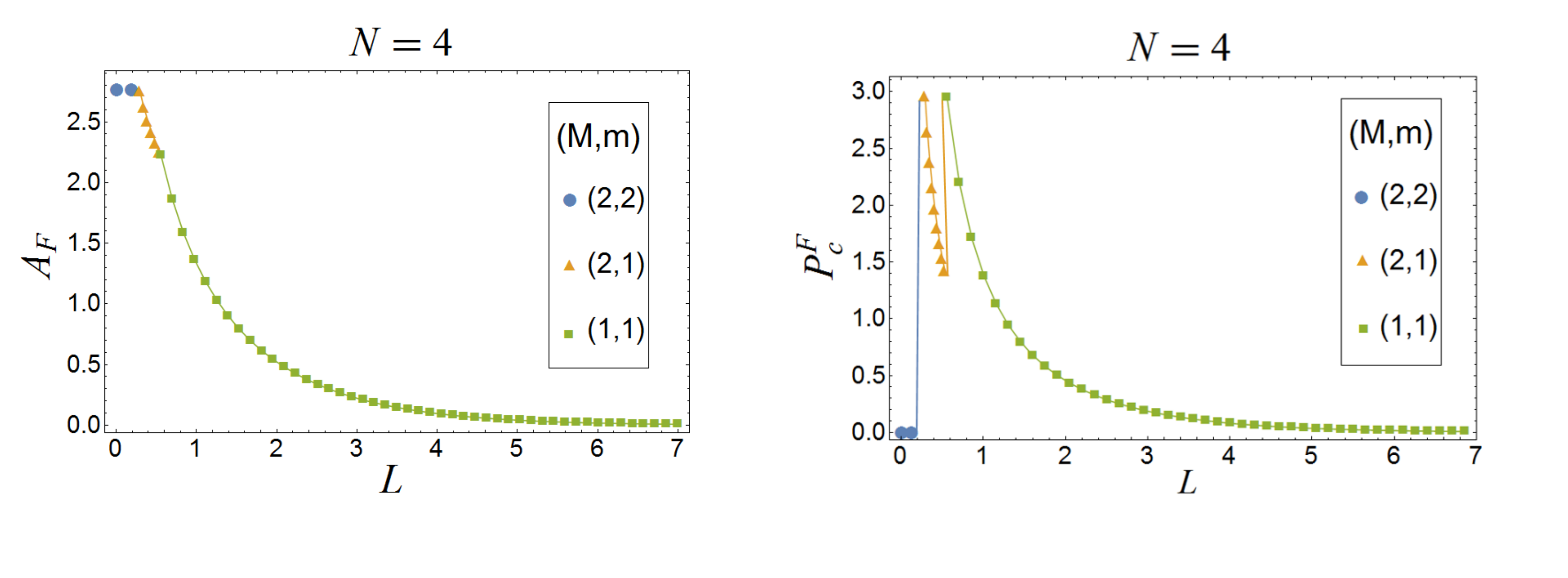}
	\caption{Fundamental Helmholtz energy $A_F$ and pressure $P_c^F$, for $N=4$ }
\label{fp}
\end{figure}

\begin{figure}[h]
	\includegraphics[width=1\linewidth]{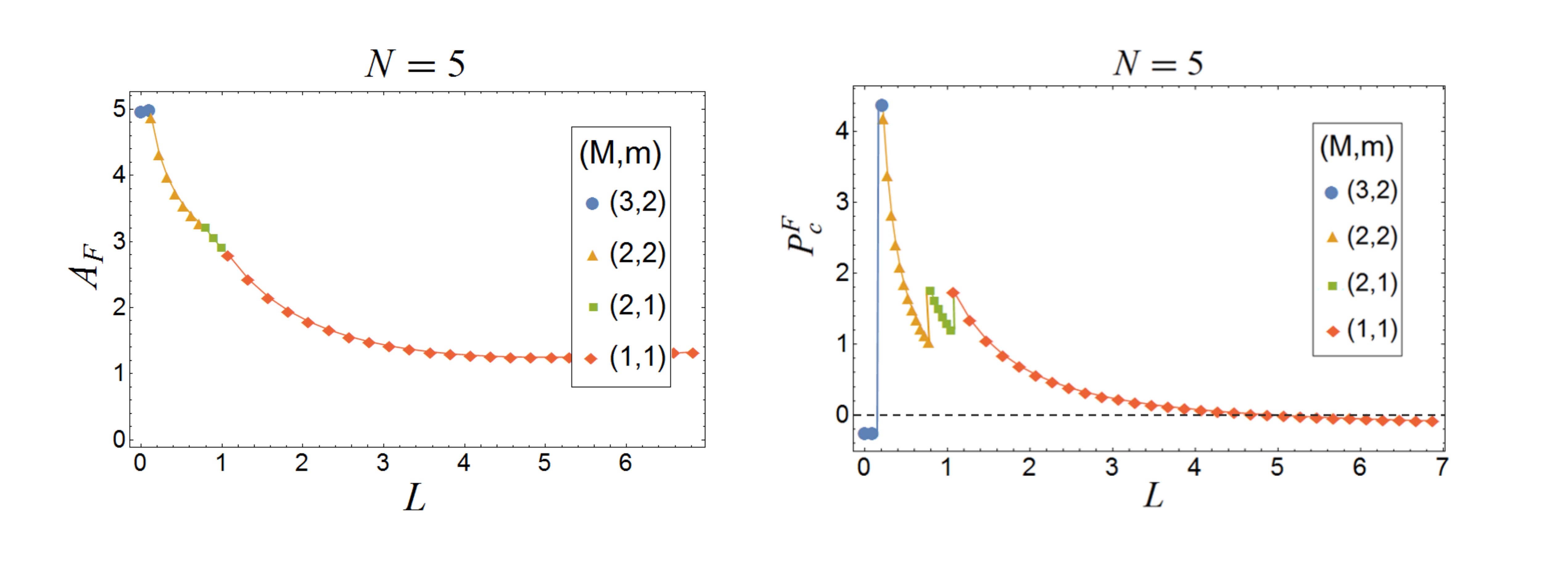}
	\caption{Fundamental Helmholtz energy $A_F$ and pressure $P_c^F$, for $N=5$ }
\label{fi}
\end{figure}

The canonical pressure is given by $P = \frac{d \ln Z_c}{dL} $. We are interested in the behavior of the fundamental (energy-minimizing) configuration.
In Figs.~\ref{fp} and~\ref{fi} one can see the essential traits for $N=4$ and $N=5$. For the even case note that the fundamental pressure is always positive while for an odd number of ions,
it can be negative. We stress again that like-charge  attraction is only possible for an odd numbers of counterions. The pressure exhibits discontinuities when a configuration transition occurs.
The Helmholtz free energy is always continuous regardless of the parity of $N$. The most notable difference between the free energies due to the parity is the asymptotic behavior
as $L\rightarrow \infty$. For $N$ even, $A$ approaches 0 as $L\to\infty$, while when $N$ is odd, $A$ tends to infinity for $L\to\infty$. This behavior was explained in section \ref{freeenergy} and it has do to with the formation of two effective opposite charges made by the colloids and the ions that screen them.

\subsection{Density Profiles}

The density profile of the counter ions is obtained by computing the following expression
\begin{equation}\label{denpro}
n(x,L) = \frac{1}{Z_c(N,L)} \sum_{k=1}^{N} \int_{x_1<\dots<x_k=x < \dots < x_N} e^{-U(x_1,\dots,x_N,L)}  \prod_{\substack{j=1 \\ j\neq k}}^{N} dx_j
\,.
\end{equation}
The sum in equation (\ref{denpro}) can be separated in three sums, one per region (in the spirit of the decomposition in Eq. (\ref{potdesc})). Due to the characteristics of the
one dimensional Coulomb potential, the integrals of each region depend only on the ions that reside in it. The parts of the integral which depend on ions that are not in the
region do cancel out, and the expression for the density profile for $x<0$ is given by
\begin{equation}\label{denleft1}
n_{\ell}(x<0,\ell) = \frac{1}{Z_L}  \sum_{k=1}^{\ell} \int_{x_1<\dots<x_k=x < \dots < x_\ell} e^{-U_L(x_1,\dots,x_\ell)}  \prod_{\substack{j=1 \\ j\neq k}}^{\ell} dx_j \qquad\qquad (x<0)
\,.
\end{equation}
This can be evaluated following similar lines as above. We introduce the change of variables
$y_j = x_j-x_k$ (for $1\leq j < k$) and the functions $\phi_k(x) = H(x) \text{exp}(k^2x)$, which yields
\begin{equation}\label{denleft2}
   \int_{x_1<\dots<x_k=x < \dots < x_\ell} e^{-U_L(x_1,\dots,x_\ell)}  \prod_{\substack{j=1 \\ j\neq k}}^{\ell} dx_j =    (-1)^{\ell-k}  \left(\Conv^{\ell}_{j=k} \phi_j(x_k)\right) \mathcal{L}\left\{ \Conv^{k-1}_{j=1} g_j(-x_1)   \right\}(0)
\,.
\end{equation}
In a second step, we have
\begin{equation}\label{denleft}
n_{\ell}(x,\ell) = \left({\ell !}\right)^2 \sum_{k=1}^{\ell} \frac{(-1)^{\ell - k}e^{k^2x}}{[(k-1)!]^2} \sum_{i=k}^{\ell} \prod_{\substack{j=k \\ j\neq i}}^{\ell} \frac{e^{i^2 x}}{i^2-j^2} \qquad\qquad (x<0)
\,.
\end{equation}
Analogously, the density profile for the region $x>L$ is obtained as:
\begin{equation}\label{denright}
n_r(x,r,L) = \left({r!}\right)^2 \sum_{k=0}^{r} \frac{e^{-(k+1)^2(x-L)}}{[k!]^2} \sum_{i=N-k+1}^{r} \;\; \prod_{\substack{j=N-k+1 \\ j\neq i}}^{r} \frac{e^{-i^2 (x-L)}}{j^2-i^2}  \qquad\qquad (x>L)
\,.
\end{equation}
For the bounded interval $x \in [0,L]$ the convolution product of the bounded partition function splits in two convolution products. This convolution products can be expressed in terms of canonical partition functions using equality (\ref{canons}):
\begin{equation}
\label{nbounded3} n_B(x,\ell,r,L) =   \frac{1}{Z_c(N,\ell,r,L) } \sum_{k=\ell+1}^{N-r}    Z_c(N,\ell,N-k+1,x) \;Z_c(N,k,r,L-x)\;      ( k!(N-k+1)!)^2
\,.
\end{equation}

\begin{figure}[h]
	\includegraphics[width=1\linewidth]{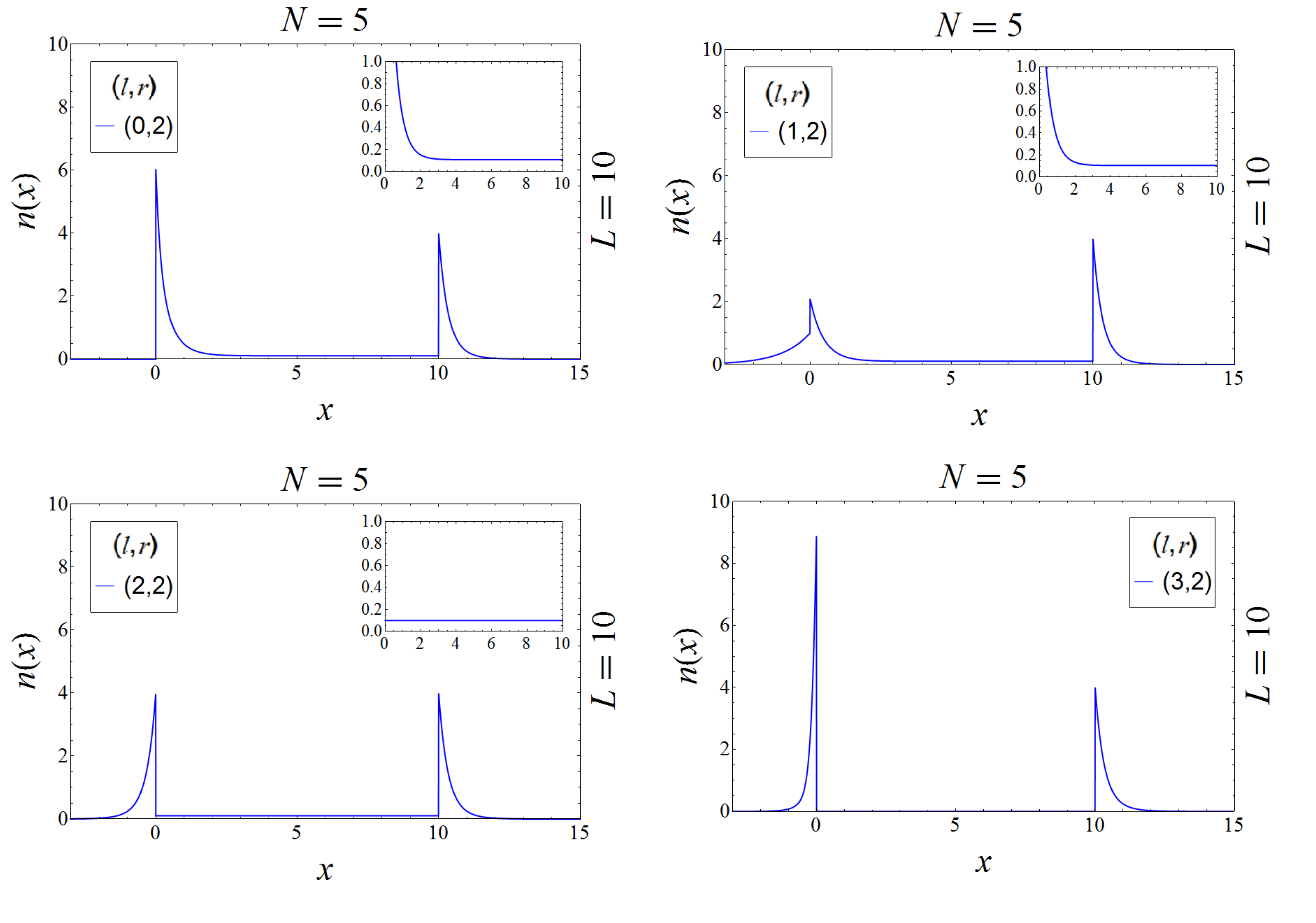}
	\caption{Density plot for fixing $r=2$ while ranging $\ell$ in $\{1,\dots,N-r\}$. }	\label{var}
\end{figure}

From the three expressions for the three regions it can be observed that for the intervals $x<0$ and $x>L$, the density is independent of $L$. Also $n_{\ell}(x,\ell)$ and  $n_{r}(x,r,L)$
do not depend on $r$ and $\ell$ respectively. All these properties
stem from the fact that in the present one dimensional setting, the presence of colloidal charges at $x=0$ and $x=L$ decouple
the corresponding half lines that each of them delimits.
In Fig.~\ref{var}, this is shown for $N=5$.
For most of the cases the density is not continuous at the colloids position. In fact for a given value of $\ell$ and $r$ it is only continuous for one value of $L$.
The values at the fixed points $x=0$ and $x=L$ follow a simple expression. For the cases $n(0^{+})$ and  $n(L^{-})$ the expressions can be obtained using the equivalent
model (\ref{equiv}), and the contact theorem (see \cite{tellez})
\begin{align}
n(0^{+}) = P + (N/2-\ell)^2 \,,\label{eq:58}\\
n(L^{-}) = P + (N/2-r)^2 \,.
\label{eq:59}
\end{align}
For the outside regions, the same equation is valid, however the pressure is zero giving the following expressions
\begin{align}
n(0^{-}) = \ell^2 \,,\\
n(L^{+}) = r^2 \,.
\end{align}
This is fully compatible with the results displayed in Fig. \ref{var}. The above results also call for a number of comments.
When no ions are bound, we necessarily have $n(0^+)= n(L^-)=0$. Eqs. \eqref{eq:58} and \eqref{eq:59} thus imply that
$(N/2-\ell)^2=(N/2-r)^2$ which is indeed true since we have $r+\ell=N$ (all ions ``out''). These equations
also immediately yield $P=0$ for $N=2n$ even, $m=M=n$ and $P=-1/4$ for $N=2n+1$ odd, $M=n+1$, $m=n$. These results have already been mentioned in section \ref{sec:optimal}.

\subsection{Charge Reversal}
Finally concerning overcharging, there is always a configuration for which this happens independently of parity or $L$. This is due to the fact that one can always pick a configuration that has $M > N/2$, which will reverse the charge of one colloid. As expected for an odd $N$ overcharging occurs for all configurations due to the fact that even for $M\leq N/2$ the charges cannot be arranged such that the screened colloids are neutral. That is, one of the colloids will have at least one more counterion in its vicinity. That and the fact that the system is neutral implies that one of the colloids will have an effective charge $q_+\geq |e|$ and the other $q_- = -q_+$.

\section{Conclusion}

We have studied a one dimensional Coulomb system, which is an
extension of the model proposed in \cite{tellez}, where the charge of
two ``colloids'' is compensated by that of an ensemble of $N$
counterions, all of the same charge.  Our goal was to identify the
fundamental (optimal) configuration of charges: how many counterions
should be confined, and therefore lying between the two colloids, and
how many should be unbound?

We found that for large separation between colloids (or small
pressures for the isobaric ensemble), the system separates in two
almost independent subsystems. These subsystems are formed by each colloid
and a screening cloud with half of the total number of
counterions. If $N$ is odd, there is additionally a single counter ion
between these subsystems. The structure of the screening cloud of each
subsystem is such that approximately half of the counterions of the
cloud are on each side of the colloid. The exact number of counterions
on each side depends on the parity of $n$ ($N=2n$ or $N=2n+1$) which
plays a crucial role in determining the subsystem structure. The
precise subsystem structure for each case is illustrated in
Figs.~\ref{fig:N26}, \ref{fig:N28}, \ref{fig:N27} and \ref{fig:N29}.

When the colloids are close (high pressures), the interaction
between counterions dominates, forcing them to escape in pairs the
bounded region. The process is symmetrical, except for the last bounded
particle in the odd case, which lacks a pair to jump with. For the
arbitrary separation of colloids, we characterize analytically the
fundamental configuration in the isobaric ensemble
(see Table~\ref{tab:transitions}).

We observed that this model allows both like-charge attraction and charge reversal, even for the fundamental configuration. As in \cite{tellez}, these phenomena were related to
the parity of the number counterions, arising due to the failure to neutralize the colloidal charges. All thermodynamic quantities for the isobaric and canonical ensembles
were obtained analytically for all the configurations. The fundamental canonical configuration was examined with  computational aid. Within the isobaric ensemble, we explained
in detail how the Gibbs energy behaves for all possible configurations, and provided the physical interpretation. We also gave asymptotic behaviors for the thermodynamic quantities in the two ensembles.

\begin{acknowledgments}
This work was supported by an ECOS Nord/COLCIENCIAS-MEN-ICETEX action
of Colombian and French cooperation (C14P01). G.~T.~acknowledges support from
Fondo de Investigaciones, Facultad de Ciencias, Universidad de los
Andes, project ``Paisaje configuracional y energ\'etico en sistemas de
Coulomb de una dimensi\'on'', 2016-2.
\end{acknowledgments}

\bibliographystyle{unsrt}
\bibliography{refs}

\end{document}